\documentclass[twoside,twocolumn,9pt]{article}
\usepackage{extsizes}
\usepackage[super,sort&compress,comma]{natbib} 
\usepackage[version=3]{mhchem}
\usepackage[left=1.5cm, right=1.5cm, top=1.785cm, bottom=2.0cm]{geometry}
\usepackage{balance}
\usepackage{mathptmx}
\usepackage{sectsty}
\usepackage{graphicx} 
\usepackage{lastpage}
\usepackage[format=plain,justification=justified,singlelinecheck=false,font={stretch=1.125,small,sf},labelfont=bf,labelsep=space]{caption}
\usepackage{float}
\usepackage{fancyhdr}
\usepackage{fnpos}
\usepackage[english]{babel}
\addto{\captionsenglish}{%
  
}
\usepackage{array}
\usepackage{droidsans}
\usepackage{charter}
\usepackage[T1]{fontenc}
\usepackage[usenames,dvipsnames]{xcolor}
\usepackage{setspace}
\usepackage[compact]{titlesec}
\usepackage{hyperref}

\usepackage{epstopdf}

\definecolor{cream}{RGB}{222,217,201}

\begin{document}

\pagestyle{fancy}
\thispagestyle{plain}
\fancypagestyle{plain}{
\renewcommand{\headrulewidth}{0pt}
}

\makeFNbottom
\makeatletter
\renewcommand\LARGE{\@setfontsize\LARGE{15pt}{17}}
\renewcommand\Large{\@setfontsize\Large{12pt}{14}}
\renewcommand\large{\@setfontsize\large{10pt}{12}}
\renewcommand\footnotesize{\@setfontsize\footnotesize{7pt}{10}}
\makeatother

\renewcommand{\thefootnote}{\fnsymbol{footnote}}
\renewcommand\footnoterule{\vspace*{1pt}%
\color{cream}\hrule width 3.5in height 0.4pt \color{black}\vspace*{5pt}} 
\setcounter{secnumdepth}{5}

\makeatletter 
\renewcommand\@biblabel[1]{#1}            
\renewcommand\@makefntext[1]%
{\noindent\makebox[0pt][r]{\@thefnmark\,}#1}
\makeatother 
\renewcommand{\figurename}{\small{Fig.}~}
\sectionfont{\sffamily\Large}
\subsectionfont{\normalsize}
\subsubsectionfont{\bf}
\setstretch{1.125} 
\setlength{\skip\footins}{0.8cm}
\setlength{\footnotesep}{0.25cm}
\setlength{\jot}{10pt}
\titlespacing*{\section}{0pt}{4pt}{4pt}
\titlespacing*{\subsection}{0pt}{15pt}{1pt}


\makeatletter 
\newlength{\figrulesep} 
\setlength{\figrulesep}{0.5\textfloatsep} 

\newcommand{\topfigrule}{\vspace*{-1pt}%
\noindent{\color{cream}\rule[-\figrulesep]{\columnwidth}{1.5pt}} }

\newcommand{\botfigrule}{\vspace*{-2pt}%
\noindent{\color{cream}\rule[\figrulesep]{\columnwidth}{1.5pt}} }

\newcommand{\dblfigrule}{\vspace*{-1pt}%
\noindent{\color{cream}\rule[-\figrulesep]{\textwidth}{1.5pt}} }

\makeatother

\twocolumn[
  \begin{@twocolumnfalse}
\vspace{1em}
\sffamily

\noindent\LARGE{\textbf{Is there a Granular Potential?}} \\

 \noindent\large{Josh M. Gramlich,\textit{$^{a}$} Mahdi Zarif,\textit{$^{b}$} and Richard K. Bowles\textit{$^{ac}$$^{\ast}$}} \\

\noindent\normalsize{Granular materials, such as sand or grain, exhibit many structural and dynamic characteristics similar to those observed in molecular systems, despite temperature playing no role in their properties.  This has led to an effort to develop a statistical mechanics for granular materials that has focused on establishing an equivalent to the microcanonical ensemble and a temperature-like thermodynamic variable. Here, we expand on these ideas by introducing a granular potential into the Edwards ensemble, as an analogue to the chemical potential, and explore its properties using a simple model of a granular system. A simple kinetic Monte Carlo simulation of the model shows the effect of mass transport leading to equilibrium and how this is connected to the redistribution of volume in the system. An exact analytical treatment of the model shows that the compactivity and the ratio of the granular potential to the compactivity determine the equilibrium between two open systems that are able to exchange volume and particles, and that mass moves from high to low values of this ratio. Analysis of the granular potential shows that adding a particle to the system increases the entropy at high compactivity, but decreases the entropy at low compactivity. Finally, we demonstrate the use of a small system thermodynamics method for the measurement of granular potential differences.} \\


 \end{@twocolumnfalse} \vspace{0.6cm}

  ]

\renewcommand*\rmdefault{bch}\normalfont\upshape
\rmfamily
\section*{}
\vspace{-1cm}


\footnotetext{\textit{$^{a}$~Department of Chemistry, University of Saskatchewan, Saskatoon, SK, S7H 0H1, Canada}}
\footnotetext{\textit{$^{b}$~Department of Physical and Computational Chemistry, Shahid Beheshti University, Tehran 19839-9411, Iran.}}
\footnotetext{\textit{$^{c}$~Centre for Quantum Topology and its Applications (quanTA), University of Saskatchewan, SK S7N 5E6, Canada. E-mail: richard.bowles@usask.ca}}



\section{Introduction}
\label{sec:intro}
Macroscopic granular materials, such as sand, grains and dry powders, are jammed systems that remain static unless subject to external forces. When shaken or tapped, the granular particles rearrange, dissipating energy through collisions, before returning to another jammed state. However, when agitated repeatedly, these systems appear to evolve to highly reproducible states in terms of their occupied volume, reminiscent of the type of reproducibility found in thermodynamics~\cite{Nowak_Density_1998,Brujic_Granular_2005}. The dynamics of the constituent particles also exhibit properties similar to those observed in molecular systems such as supercooled liquids and glasses~\cite{Philippe_Compaction_2002,Reis_Caging_2007}, or complex fluids~\cite{Kou_Granular_2017}. This has led to the suggestion that the properties of granular materials could be described by an appropriate form of statistical mechanics. As a starting point, Edwards~\cite{Edwards_Theory_1989,Mehta_Statistical_1989,Edwards_Statisitcal_1998,Blumenfeld_Geometric_2006} proposed the equivalent to the micro-canonical ensemble by postulating all jammed packings with the same volume are equiprobable, which then defines the entropy as the number of jammed states and the compactivity, a measure of the capacity of the system to be compressed, as the granular equivalent to the thermal temperature. 

A considerable amount of theoretical, computational and experimental work has followed and different aspects of the effort to develop a clearer picture of how statistical mechanics can be used to describe granular materials have been extensively reviewed~\cite{Chakraborty_statistical_2010,Chakraborty_Statisitcal_2015,Baule_Edwards_2018}. Studies of the Edwards volume ensemble~\cite{Aste_Volume_2006,Briscoe_Entropy_2008,McNamara_Measurement_2009}, using different volume measures, have been successful in describing the properties of a number of granular systems. This includes establishing a phase diagram for jammed matter that spans the random loose packing to random close packing limits, where the compactivity, $X\rightarrow \infty$ and $X\rightarrow 0$, respectively, and accounts for effects due to particle friction~\cite{Song_Phase_2008,wang_Jamming_2010}. However, direct tests of the Edwards hypothesis have generally been less conclusive. For example, studies of small systems of binary discs, where the jammed states can be studied through exhaustive simulation~\cite{Gao_Experimental_2009} have shown that packings with the same volume are not necessarily equally probable. Recent simulations that measured the configurational basin volumes associated with packings of polydisperse soft discs found that the Edwards hypothesis was correct at the jamming point, where the configurational entropy is maximal, but was not generally true at other densities~\cite{Martiniani_Numerical_2017}.

Experiments~\cite{Puckett_Equilibrating_2013} have also found that the compactivity does not equilibrate between systems in contact, suggesting the stress ensembles~\cite{Blumenfeld_Granular_2009,Chakroborty_Statistical_2010,Chakraborty_Statisitcal_2015}, force-network ensembles~\cite{Tighe_Force_2010}, or ensembles that include both volume and stress microstates~\cite{Blumenfeld_Inter_2012,Wang_Edwards_2012}, may provide the basis for understanding granular systems. Despite this progress, the identification of a granular temperature, and the validity of a zeroth law for granular thermodynamics, remains an open question, with recent experiments providing contrasting perspectives. Bililign et al~\cite{Bililign_Protocol_2019} demonstrate that the force tile-area, or keramicity, provides a temperature-like quantity in the force-moment ensemble for a two-dimensional granular system in a work that also highlights how protocol dependence complicates the search for a statistical mechanics of granular materials. Meanwhile, Yuan et al~\cite{Yuan_Experimental_2021,Zeng_Equivalence_2022} find that the original Edwards volume ensemble and the compactivity are sufficient to establish a zeroth law of granular thermodynamics in a system of tapped granular packings, including systems with friction.

 As the field of granular statistical mechanics develops, it is worth starting to ask the question of whether there are granular analogies to the other thermodynamic potentials used in chemical systems. In particular, the chemical potential plays a central role in our understanding of equilibrium and mass transport in molecular systems. It also provides the basis for additional ensembles in statistical mechanics, as well as a variety of molecular simulation techniques, which suggests an equivalent quantity could be useful in the study of granular systems. The concept of a non-equilibrium chemical potential has been explored in the context of driven non-equilibrium steady state systems~\cite{Bertain_Definition_2006,Dickman_Inconsistencies_2014,Guioth_Nonequilibrium_2021}, such as in active matter~\cite{Guioth_Nonequilibrium_2021,Paliwal_Chemical_2018}. Less is known with respect to the role a chemical potential plays in dense granular systems that are driven, but move between static, jammed states. A granular chemical potential has been used to study the possibility of a first order phase transition between disordered and crystalline jammed packings in hard spheres~\cite{Jin_First_2010} and to motivate the use of partial molar volumes in the analysis of bi-disperse granular packings~\cite{Chang_Compaction_2022}, but more direct investigations of the granular chemical potential are not available. 

The goal of this work is to explore the role of the granular chemical potential, or granular potential to distinguish it from its chemical equivalent, in equilibrium and its properties as a function of its thermodynamic parameters in a simple system granular model. We also demonstrate how the granular potential can be measured within the Edwards ensemble. The remainder of the paper is organized as follows: Section~\ref{sec:gp} provides a thermodynamic definition of the granular potential and its role in defining equilibrium in the Edwards ensemble. Section~\ref{sec:app} describes the application of the granular potential to a simple model of a granular material, including a kinetic Monte Carlo simulation showing mass transport leading to equilibrium, and two analytical examples of equilibrium in systems able to exchange mass and volume. In Section~\ref{sec:mea}, we show how differences in the granular potential can be measured using a small system approach. Our discussion and conclusions are contained in Sections~\ref{sec:dis} and ~\ref{sec:con}, respectively.


\section{A Granular Potential}
\label{sec:gp}
The Edwards ensemble~\cite{Edwards_Theory_1989,Edwards_Statisitcal_1998} provides a starting point for a statistical mechanics of granular systems by defining an ensemble where all the jammed states of a system with the same volume, $V$, and a fixed number of particles, $N$, are equally probable. The probability of finding a given jammed state is then,
\begin{equation}
P_i=e^{-S/\lambda}\delta(V-W_i)\Theta\mbox{,}\\
\label{eq:pi}
\end{equation}
where $W_i$ is the Hamiltonian-type volume function for the configuration, $S$ is the entropy, $\lambda$ is the analog to the Boltzmann constant,
\begin{equation}
\Theta=
\begin{cases}
1&\textit{if collectively jammed}\\
0&\textit{otherwise}
\end{cases}
\mbox{,}
\label{eq:cj}
\end{equation}
is the collective jamming condition and the normalization of the probability is given by,
\begin{equation}
e^{S/\lambda}=\int\delta(V-W_i)\Theta d(\textit{all degrees of freedom})\mbox{.}\\
\label{eq:norm}
\end{equation}
Integrating Eq.~\ref{eq:norm} and taking the natural log yields,
\begin{equation}
S/\lambda=\ln\Omega(V,N)\mbox{,}\\
\label{eq:entropy}
\end{equation}
where $\Omega(V,N)$ is the number of collectively jammed configurations.

In conventional thermal statistical mechanics, the partition function is connected to the other thermodynamic parameters through the entropy and the first law, but there is no clear definition of heat and work for granular systems. Instead, we can write an analogue to the thermodynamic fundamental equation, where the volume replaces the role of the energy, and introduce new granular thermodynamic potentials that affect variations of the volume, to give for example, 
\begin{equation}
dV=XdS+\gamma dN\mbox{.}\\
\label{eq:fun}
\end{equation}
Equation~\ref{eq:fun} defines Edwards compactivity for a system with fixed $N$,
\begin{equation}
X=\left(\frac{\partial V}{\partial S}\right)_N\mbox{,}\\
\label{eq:comp}
\end{equation}
as a granular equivalent to the thermal temperature that measures the capacity of the system to be compressed. 

The second term on the right of the equality in Eq.~\ref{eq:fun} introduces the granular potential, $\gamma$, as the intensive conjugate variable to $N$ that captures the effect of adding a particle to the system, as an analogy to the chemical potential. We can also use the fundamental equation to establish the key characteristics of the granular potential. For example, at constant volume, $dV=0$, so Eq.~\ref{eq:fun} rearranges to give,
\begin{equation}
\left(\frac{\partial S}{\partial N}\right)_V=-\frac{\gamma}{X}\mbox{.}\\
\label{eq:dsdn}
\end{equation}
At constant $V$, the insertion of a particle has two entropic effects. The entropy is an extensive quantity, so adding a particle necessarily increases $S$. However, the added particle also occupies volume, which increases the density of the system, reducing the number of accessible jammed states. The nature of the two contributions to the granular potential can be seen clearly by taking the derivative of Eq.~\ref{eq:fun} with respect to $N$ at constant $X$, to obtain,
\begin{equation}
\frac{\gamma}{X}=\frac{1}{X}\left(\frac{\partial V}{\partial N}\right)_X-\left(\frac{\partial S}{\partial N}\right)_X\mbox{,}\\
\label{eq:gox}
\end{equation}
where $(\partial V/\partial N)=V/N$ and  $(\partial S/\partial N)=S/N$ are the partial molar volume and partial molar entropy, respectively, which become their molar quantities for a single component system.

The conditions for equilibrium in a conventional thermodynamic system are obtained by maximizing the total entropy, $S=S_1+S_2$, of an isolated composite system, composed of two subsystems in contact with each other, that are able to exchange energy, mass and volume, subject to conservation conditions. We can do the same here. Writing Eq.~\ref{eq:fun} for subsystems 1 and 2 and using the conservation of mass, $dN=dn_1+dn_2=0$, and volume, $dV=dV_1+dV_2$, to obtain the variation of the total entropy, 
\begin{equation}
\begin{split}
dS&=dS_1+dS_2=0\\
& =\left(\frac{1}{X_1}-\frac{1}{X_2}\right)dV_1-\left(\frac{\gamma_1}{X_1}-\frac{\gamma_2}{X_2}\right)dn_1\mbox{.}
\end{split}
\label{eq:dseq}
\end{equation}
Both $n_1$ and $V_1$ are independent, so Eq.~\ref{eq:dseq} is only satisfied if the coefficients are zero. At constant $n_1$ the systems are in equilibrium when $X_1=X_2$, at constant $V$ the systems are at equilibrium when $\gamma_1/X_1=\gamma_2/X_2$ and both equalities are required for general equilibrium.

\section{Application to a Simple Model}
\label{sec:app}

\begin{figure}[]
\centering
\includegraphics[width=8.0cm]{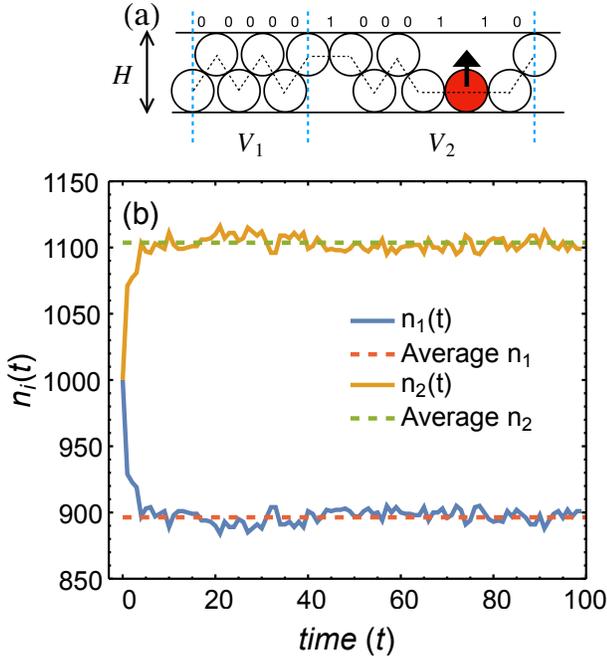}
\caption{(a) Granular model showing the dense (0) and defect (1) states. Neighbouring defects (11) allow the shaded (red) particle to become unjammed. The vertical dashed lines denote the subsystems with volumes, $V_1$ and $V_2$. (b) The number of particles in each subsystem as a function of time (solid lines) compared to their averages (dashed lines). } 
\label{fig:traj}
\end{figure}

To obtain a concrete understanding of the role the granular potential plays in equilibrium we study a simple model consisting of two-dimensional (2D) hard discs, with diameter $\sigma$, confined between hard walls separated by a distance $1< H/\sigma<1+\sqrt{3/4}$ (see Fig~\ref{fig:traj}(a)). The model has been used previously to study the compactivity in granular systems~\cite{Bowles_Edwards_2011}, as well as glassy relaxation~\cite{Bowles_Landscapes_2006,Yamchi_Fragile_2012,Ashwin_Inherent_2013,Godfrey_static_2014,Yamchi_Inherent_2015,Godfrey_Absence_2018}, because all of the jammed states can be fully characterized. Briefly, in 2D, a particle is locally jammed if it has three rigid contacts, not all located in the same semicircle, but the local jamming of all the particles is not sufficient for collective jamming~\cite{Torquato_Multiplicity_2001} as the concerted motion of groups of particles can lead to unjamming. The restriction on $H/\sigma$ prevents such rearrangements. It also ensures that discs can only touch their nearest neighbour on each side and the wall. As a result, there are only two local jamming environments, a dense structure with a volume $v_0=H\sqrt{(2\sigma-H)H}$, formed when two discs are in contact across the channel, and a defect state with volume $v_1=H\sigma$, formed when two discs are in contact along the channel. The total volume per particle of a jammed state is then given by,
\begin{equation}
V/N=(1-\theta)v_0+\theta v_1\mbox{,}\\
\label{eq:von}
\end{equation}
where $\theta=N_1/(N_0+N_1)$ is the fraction of defects in the system, with $N_0$ and $N_1$ being the number of dense and defect states in the packing respectively, and $N_1+N_0=N$. The whole system remains collectively jammed as long as there are no neighbouring defects (see Fig.~\ref{fig:traj}(a), adopting the most dense jammed state as $\theta\rightarrow 0$ and the least dense jammed state as $\theta\rightarrow 0.5$. The entropy of the system is also just a function of defect fraction~\cite{Bowles_Landscapes_2006},
\begin{equation}
S(\theta)/\lambda N=(1-\theta)\ln(1-\theta)-(1-2\theta)\ln(1-2\theta)-\theta\ln\theta\mbox{,}\\
\label{eq:son}
\end{equation}
that exhibits a maximum when $\theta=1/2-\sqrt{5}/10$.
All the packings of the system are isostatic and the model satisfies the Edwards postulate of equiprobable states. In the limit $H/\sigma\rightarrow1$, the system approaches the strict one-dimensional hard rod system and the distribution of jammed states collapses to a single jammed state. When $H/\sigma > 1+\sqrt{3/4}$ the discs can touch their next nearest neighbours which leads to a more complex jamming landscape~\cite{Ashwin_Complete_2009}, including unusual asymptotically crystalline states~\cite{Zhang_Marginally_2020}, where many of these packings are not isostatic and it is not known if the system is described by the Edwards ensemble.

Discrete element simulations of the model~\cite{Irastorza_Exact_2013}, for both frictionless and frictional particles, show there are important differences between the idealized model and a more realistic representation of a granular system, with the simulations being unable to sample states associated with $X\rightarrow\infty$ and $X\rightarrow 0$. At small $X$, it is likely the tapping is non-ergodic, so defects are slow to diffuse preventing the system from reaching equilibrium, but at high $X$ it is possible that the nature of the dynamics influences how the states are sampled, which highlights the important role protocol plays in determining the properties of real granular materials. 
Nevertheless, the model correctly predicts many properties, including the states sampled and the properties of the volume fluctuations, suggesting the underlying Edwards ensemble still provides a starting point for describing the statistical mechanics of the system.

The process of mass transport, leading to equilibrium, in our granular system can best be demonstrated using a simple simulation.  We consider a system of $N=2000$ discs and $H/\sigma=1.86$, divided into two subsystems with volumes $V_1/\sigma^2=949.15$ and $V_2/\sigma^2=1167.75$, with $V=V_1+V_2$, both initially containing $n_1(t=0)=n_2(t=0)=1000$ particles. The particles in subsystem 1 are initially arranged in the most dense state, with no defects ($\theta_1=0$), while we randomly place defects in subsystem 2 such that there are no neighbouring defects and $\theta_2=0.24$. The location of the boundaries separating the two systems is fixed but particles can pass through. Granular systems need dynamics that take them from jammed state to jammed state so they can sample the ensemble of states. Experimentally, this is usually achieved by shaking or tapping the system, which provides the particles with kinetic energy, causing the system to expand and opening up pathways in configurational space that allow the system to rearrange. As a consequence, the system naturally samples the canonical ensemble, exchanging volume with the surrounding. It is important to note that the ensemble is made up of the jammed states, but the system moves through unjammed configurations as it jumps from one jammed state to another in a way that is determined by the particular dynamics or protocol of the system. This is likely to influence the probability with which the states are sampled. To simulate the equilibration of the isolated system, where $V$ is fixed, we take advantage of our knowledge of the complete set of jammed states and use a simple kinetic Monte Carlo (KMC) scheme that takes the system directly between jammed states with moves that satisfy detailed balance (see Appendix for details). The MC moves employed by our method represent local exchanges, where particles swap environments with their neighbours so the defect states essentially diffuse through the system, but non-local events also occur when two defects diffuse together and are eliminated. A defect pair must then be generated somewhere in the system in order to maintain the fixed $V$. Our method allows the system to explore configuration space and evolve towards equilibrium without any dynamical bias and avoids any protocol dependence in the sampling of the ensemble. As described, our composite system represents an isolated granular system with fixed total volume and total number of particles. The two subsystems also have fixed volumes but are able to exchange particles, subject to the constraint $n_1+n_2=N$. 

As described, our composite system represents an isolated granular system with fixed total volume and total number of particles. The two subsystems also have fixed volumes but are able to exchange particles, subject to the constraint $n_1+n_2=N$. 
 The simulation is run for $20000$ time steps and average quantities are calculated over the last 18000 time steps.
Figure~\ref{fig:traj}(b) shows the evolution of $n_i(t)$, the number of discs in each subsystem, as a function of time. Both subsystems start with the same number of particles, but discs rapidly move from subsystem 1, where the density is high, into subsystem 2, until both $n_i(t)$ reach a plateau after approximately 20 time steps and fluctuate around what is clearly their equilibrium values, $\left<n_1\right>=896\pm5$ and $\left<n_2\right>=1103\pm5$. Here, the error is the standard deviation of the data. The degree of confinement in the model explicitly prevents the particles from passing each other so they cannot diffuse. Instead, the defects diffuse through the system, with defect pairs sometimes annihilating and then spontaneously reforming in another part of the system in order to keep the total defect fraction constant and the system jammed. This leads to a redistribution of the volume and as the defects diffuse into $V_1$, effectively expanding average volume per particle in this region, some particles are pushed from $V_1$ into $V_2$, until equilibrium is obtained. In a molecular system, the transport of material between subsystems would be driven by differences in the chemical potential and the system would come to equilibrium when the chemical potential is the same in each system. The mass transport observed in our simulation suggests that a granular potential may serve a similar role.

\begin{figure}[]
\includegraphics[]{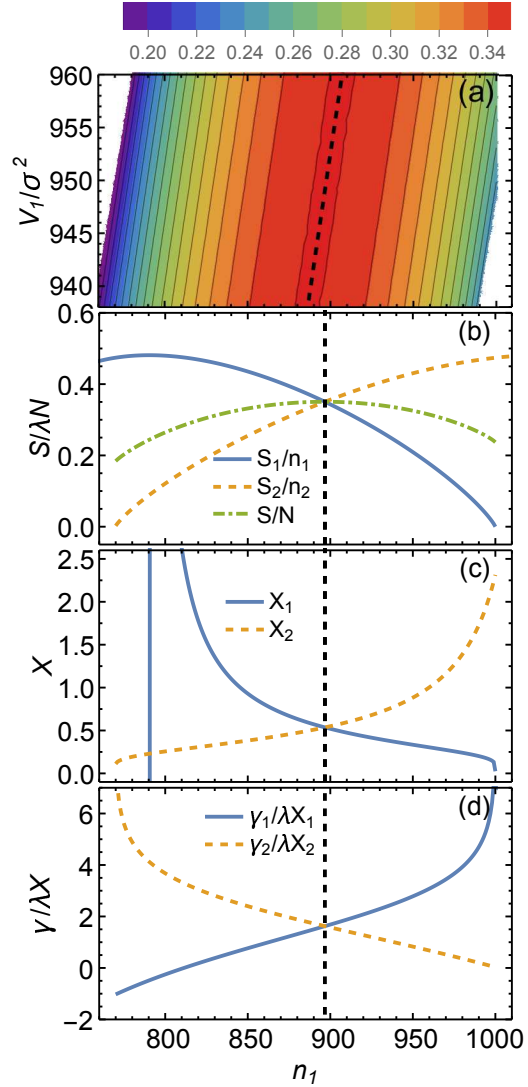}
\caption{ (a) Contour plot of total entropy, $S$, as a function of $n_1$ and $V_1$, with $V/\sigma^2=2116.9$, $N=2000$ and $H_1/\sigma=H_2/\sigma= 1.86$. Dashed line indicates a ridge of equivalent maxima. (b) Entropy, (c) compactivity and (d) granular potential as a function of $n_1$ for the composite system with fixed $V_1/\sigma^2=949.15$ and $V_2/\sigma^2=1167.75$. Vertical black dashed lines denote the equilibrium value of $n_1$.} 
\label{fig:equil1}
\end{figure}

The granular properties of our composite system can be obtained analytically as a function $n_1$, using,  Eqs.~\ref{eq:von}, ~\ref{eq:son}, the conservation of mass and volume, along with Eqs.~\ref{eq:comp} and \ref{eq:gox}, where we have also used $(\partial V/\partial S)_N= (\partial V/\partial \theta)_N(\partial \theta/\partial S)_N$. Figure~\ref{fig:equil1}(a) shows the total entropy for the composite system exhibits a line of equivalent maxima  corresponding to the equilibrium defined by Eq.~\ref{eq:dseq}. Figure~\ref{fig:equil1}(b) shows the entropy per particle for a system with the volume $V_1$ corresponding to our simulation. The upper bound to $n_1$ occurs when subsystem 1 contains enough particles to be in the most dense jammed state, where $\theta_1=0$ and $S_1=0$. The lower bound to $n_1$ occurs when enough particles have moved into subsystem 2 to cause it to be in the most dense jammed state, where $\theta_2=0$ and $S_2=0$. The equilibrium value for $n_1=896.7$, obtained as the maximum in entropy of the isolated composite system, $S=S_1+S_2$, is the same as that obtained in the simulation. The compactivity in subsystem 1 (Fig.~\ref{fig:equil1}(c)) increases as $n_1$ decreases, and actually diverges when $S_1$ goes through its maximum. The divergence point appears as the vertical line in $X_1$. At smaller $n_1$, the number of jammed states in the system begins to decrease and  the compactivity becomes negative, which is analogous to the appearance of negative temperatures in thermal systems with similar distributions in their density of states. At the same time, $X_2$ decreases with decreasing $n_1$ and $X_1=X_2$ at equilibrium. Figure~\ref{fig:equil1}(d) shows $\gamma_1/X_1$ decreases, and $\gamma_2/X_2$ increases, as particles move from subsystem 1 to subsystem 2, and $\gamma_1/X_1=\gamma_2/X_2$ at equilibrium. This confirms that mass, in a granular system, moves from regions of high $\gamma/X$ to regions of low $\gamma/X$. If we choose a different $V_1$, the maximum occurs at a different $n_1$ along the dashed line, but the equilibrium values of $X$ and $\gamma/X$ remain the same because the compactivity and granular potential should be uniform throughout the system and independent of where we choose the dividing surface for subsystems.

\begin{figure}[]
\centering
\includegraphics[]{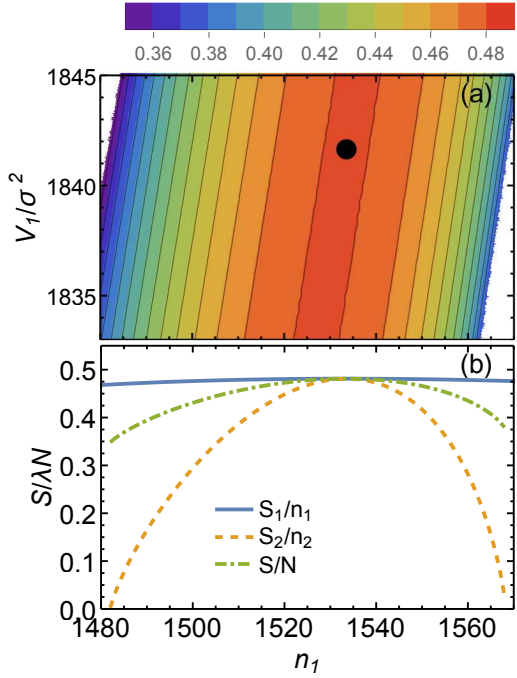}
\caption{(a) Contour plot of total entropy, $S$, as a function of $n_1$ and $V_1$, with $V/\sigma^2=2470.58$, $N=2000$, $H_1/\sigma=1.86$ and $H_2/\sigma=1.7$. Black point indicates maximum. (b) Entropy for system with $V_1/\sigma^2=1841.64$ corresponding to equilibrium.}
\label{fig:equil2}
\end{figure}

\begin{figure}[]
\centering
\includegraphics[]{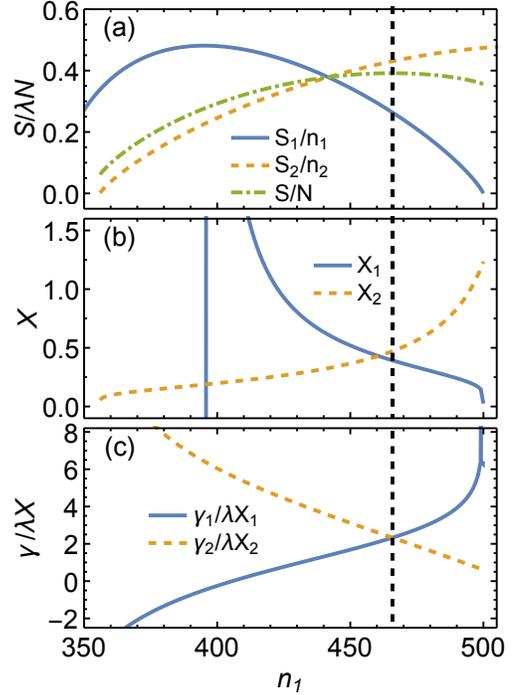}
\caption{(a) Entropy, (b) compactivity and (c) granular potential as a function of $n_1$ for a composite system with, $N=2000$, $H_1/\sigma=1.86$, $H_2/\sigma= 1.70$, $V_1/\sigma^2=465.83$ and $V_2/\sigma^2=1996.01$. Vertical black dashed line denotes the equilibrium value of $n_1$.} 
\label{fig:equil2b}
\end{figure}

                                                                                                                                                                                                                                                                                                                                                                                                                                                                                                                                                                                                                                                                                                                                                                                                                                                                                                    To test the generality of our result, we build a second composite system with $H_1/\sigma=1.86$ and $H_2/\sigma=1.70$, so the particles in the different height channels occupy different volumes when they are in the two allowed local jamming environments. This example is equivalent to allowing molecular exchange between two systems with distinct chemical environments, where we would still expect the chemical potential in the systems to be equal at equilibrium. Figure~\ref{fig:equil2}(a) shows that the composite system exhibits a single maximum, despite the near parallel appearance of the contour lines, corresponding to the coexistence between the two subsystems. However, the equilibrium is located at the point where both subsystems have reached their maximum in entropy (See Fig.~\ref{fig:equil2}(b)) so we find $X_1=X_2=\infty$ and $\gamma_1/X_1=\gamma_2/X_2\approx-0.4812$, which is minus the entropy per particle at the maximum of the entropy maximum and is consistent with the expectations of Eq.~\ref{eq:gox} at $X=\infty$. We can also consider equilibrium in the composite system where volumes $V_1$ and $V_2$ are fixed. For example, Fig.~\ref{fig:equil2b} shows the thermodynamics of the composite system with $V_1$ fixed away from the global equilibrium. At the maximum $S\approx0.3916$, which is lower than the global maximum as expected, and $S_1\neq S_2$. As a result, we still find that $\gamma_1/X_1=\gamma_2/X_2$, but since $dV_1=0$, the equality of compactivities in Eq.~\ref{eq:dseq} is no longer required, and $X_1\neq X_2$.  

\begin{figure}[t]
\centering
\includegraphics[]{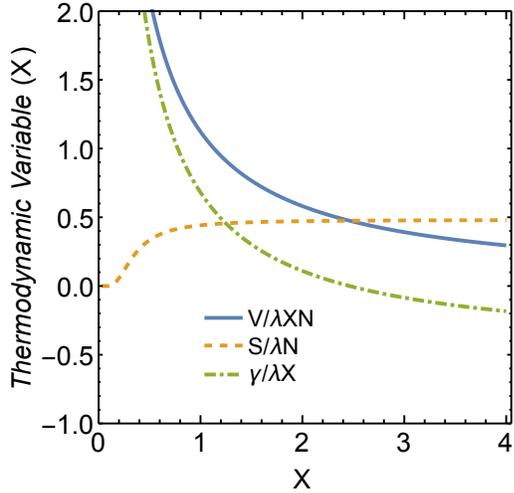}
\caption{Thermodynamic contributions to  $\gamma/X$ from Eq.~\ref{eq:gox} as function of compactivity, for a system  $H/\sigma=1.86$.}
\label{fig:muox}
\end{figure}

The granular potential gives us insight into the effects of adding a particle to a jammed system, but it is important to recognize that particle insertion in a granular system is fundamentally different from that in a chemical system. The particles in the initial jammed state must rearrange to accommodate the new particle and ensure the final state is also jammed, which is not the case, for example, in the Widom insertion method~\cite{Frenkel_Book} used to calculate the chemical potential. This has consequences for the effects of adding a particle to the system. Figure~\ref{fig:muox} plots $\gamma/X$ as a function of the compactivity, along with its two thermodynamic contributions identified in Eq.~\ref{eq:gox}.  At high $X$, there are a large number of expanded jammed states, so the entropic term makes a large contribution, while the volume term, scaled by the compactivity, is small. The resulting value for $\gamma/X$ (with $X>0$) is negative, which implies that adding a particle to the system at constant volume increases the entropy (Eq.~\ref{eq:dsdn}). However, as $X$ decreases, the volume term begins to increase because jammed states are more compact and the system needs to eliminate more defect states in order to accommodate the new particle. At the same time $S/N$ is decreasing, so eventually $\gamma/X$ changes sign, indicating the addition of a particle to the system will lead to a decrease in the entropy.

\section{Measuring the Granular Potential}
\label{sec:mea}
The chemical potential of a thermal system can be measured using a variety of methods in molecular simulation~\cite{Frenkel_Book}. Many of these
compare the properties of ensembles containing $N$ and $N+1$ particles, but the global nature of the jamming constraint makes this difficult and time consuming to evaluate. 
To overcome these difficulties, we propose using the small system ensemble method recently developed by Br\r{a}ten {\it et al.}~\cite{Braten_Chemical_2021}. The method measures the difference in chemical potential between two systems by embedding a small subsystem of volume $v$ in each of the bulk systems and using the overlap distribution method~\cite{Frenkel_Book,Bennett_Efficient_1976,Shirts_Equilibrium_2003,Shirts_Simple_2013} (ODM) to extract the chemical potential difference from their respective $n$-particle probability distributions. This yields a size dependent chemical potential for the small subsystem, where small system boundary effects are important, so the true chemical potential is obtained by making measurements for different sized subsystems and extrapolating to the thermodynamic limit.

To adapt this approach to a granular system described by the Edwards ensemble, we start by considering the properties of a small system of fixed volume $v$ embedded inside a bulk system that is large enough to act as a particle reservoir. The bulk system samples jammed states according to the canonical ensemble at a given $X$ and so fluctuates in volume $V$. This allows the particles to move as the system hops between jammed states, which causes the number of particles contained within the subsystem $v$ to fluctuate. Since the bulk system states are jammed, the particles inside $v$ are also jammed. However, the finite size of the subsystem means there are interfacial effects where particles, and the jamming constraints that fix the particles in place, cross the boundary. For a subsystem system with a given boundary, we can then identify three types of particles, those completely contained within $v$, those that have particle centres contained within $v$ but have excluded volume that crosses the boundary, and those that have particle centres outside the subsystem boundary but have excluded volumes that extend inside. We can count the number of particles in the cell, $n$, by simply considering the number of particles completely inside the cell, $n_v$, i.e. $n=n_v$. This would provide a strict lower bound on $n$ that will approach the correct thermodynamic value as $v$ increases and surface effects become less important. Alternatively, we can average the effects of the two types of boundary particles, giving
\begin{equation}
n=n_v+(1/2)(n_{s1}+n_{s2})\approx n_v+n_{s1}\mbox{,}\\
\label{eq:ns}
\end{equation}
where $n_{s1}$ and  $n_{s2}$ are the number of boundary particles with their centres inside and outside the subsystem, respectively, and the approximation on the right assumes the boundary particles location relative to the boundary is random. In the thermodynamic limit, $n_{s1}$ again becomes insignificant, i.e. $n_{s1}/v\rightarrow 0$ as $v\rightarrow \infty$.

The small subsystem represents a grand microcanonical ensemble (see Appendix~\ref{sec:gmc}) with a fixed volume $v$, immersed in a particle reservoir that fixes $\gamma/\lambda X$ in the subsystem. The probability of observing $n$ particles in $v$ is then given by,
\begin{equation}
P(n)=\frac{\Omega(n,v)\exp\left[\gamma n/\lambda X \right]}{Z_{\gamma/\lambda X,v}}\mbox{,}\\
\label{eq:pn}
\end{equation}
where $Z_{\gamma/\lambda X,v}$ is the grand microcanonical partition function given by Eq.~\ref{eq:z}. The ODM method compares the $n$-probability distributions at two different thermodynamics states, $\gamma_1/\lambda X_1$ and $\gamma_2/\lambda X_2$. Taking the natural log of the ratio of the probabilities for the two subsystems yields,
\begin{equation}
\begin{split}
\ln\frac{P_2(n)}{P_1(n)}&=(\gamma_2/\lambda X_2 -\gamma_1/\lambda X_1) n +(1/\lambda X_2 -1/\lambda X_1) v\\
&=a_1 n -a_0
\end{split}
\label{eq:a1a0}
\end{equation}
where we have used Eq.~\ref{eq:zv} to replace terms involving $\ln Z_{\gamma/\lambda X,v}$, and the fact that $\Omega(n,v)$ is the same in both subsystems, to obtain a linear relationship in $n$. Equation~\ref{eq:a1a0} gives the difference, $(\gamma_2/\lambda X_2 -\gamma_1/\lambda X_1)$ as the slope, $a_1$, and $(1/\lambda X_2 -1/\lambda X_1)v$ as intercept, $a_0$. Notice we only obtain differences in the thermodynamic quantities and there are only two independent quantities. However, the compactivities may well be known by other means leaving only one difference to be found.

\begin{figure}[t]
\centering
\includegraphics[]{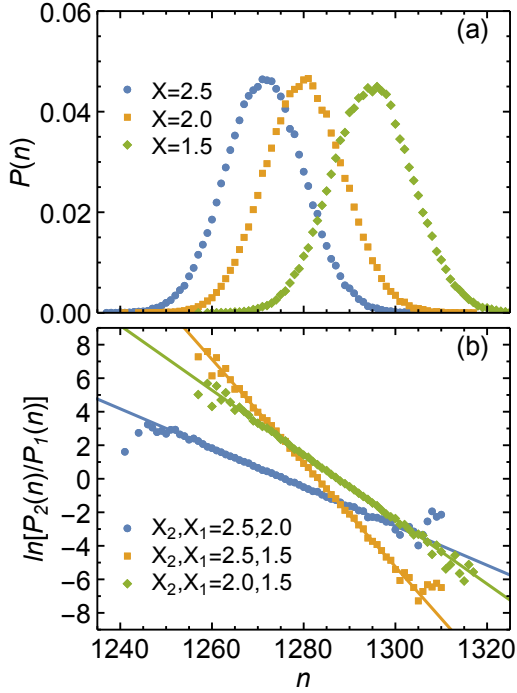}
\caption{(a) Small system $P(n)$ as a function of $n$ obtained from canonical simulations at different $X$ for $H/\sigma=1.86$, $N=5000$ and $v/\sigma^2=1488$. (b) $\ln [P_2(n)/P_1(n)]$ as a function of $n$ obtained from simulation (points) compared to exact results of the model used in Eq.~\ref{eq:a1a0} (solid lines).}
\label{fig:pn}
\end{figure}

To obtain $P(n)$ we perform Monte Carlo canonical ensemble simulations (See Appendix~\ref{sec:ces} for details) in a system with $N=5000$, at different $X$, and $H/\sigma=1.86$. After the system reaches equilibrium at each compactivity, we sample configurations separated by a minimum of $20$ MC cycles, by randomly selecting a point along the length of the system to act as the center of the subsystem and counting $n$, for a series of $v/\sigma^2=37.2, 74.4, 186.0, 279.0, 372.0, 744.0, 1488.0, 1860.0$.  The boundary of the subsystem crosses the whole channel so the surface area, $2 H$, remains fixed for all $v$. The probability distributions are calculated by sampling $1.25\times10^5$ configurations and constructing the histogram of each $n$. Figure~\ref{fig:pn}(a) shows the overlapping distributions obtained for a subsystem at three different $X$. Comparing each pair of thermodynamic states using Eq.~\Ref{eq:a1a0} (see Fig~\ref{fig:pn}) yields the expected linear behaviour, consistent with the exact results.

\begin{figure}[t]
\centering
\includegraphics[]{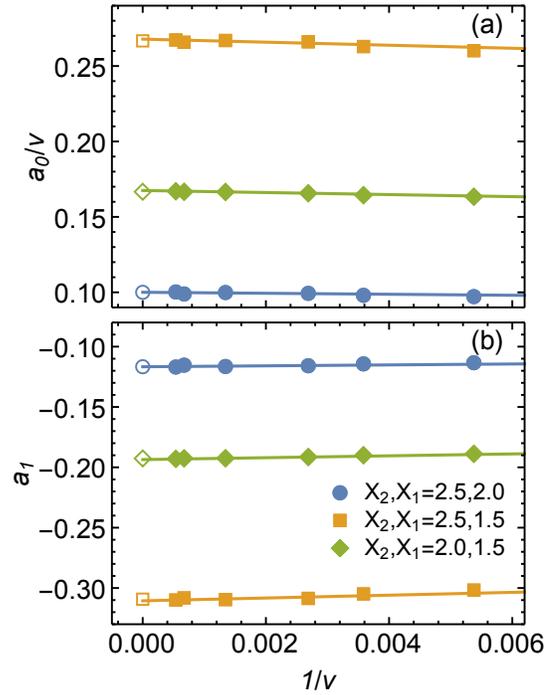}
\caption{Linear parameters (a) $a_0$ and (b) $a_1$ obtained using maximum likelihood (filled symbols) as a function of $1/v$. Errors estimated as two standard deviations of 200 bootstrap samples of the maximum likelihood are smaller than the symbols. Solid lines represent linear fits to the data and exact results  (open symbols) are located at $1/v=0$.}
\label{fig:like}
\end{figure}

In principle, the linear coefficients can be extracted by fitting the histogram data, but the values of $a_1$ and $a_0$ can be more accurately determined using a maximum likelihood approach~\cite{Shirts_Equilibrium_2003,Shirts_Simple_2013}, which for the current system is given by,
\begin{equation}
\ln L(a_1,a_0|\mbox{data})=\sum_1^{k}\ln f(-a_0-a_1 n)
+ \sum_1^{k}\ln f(a_0+a_1 n)
\end{equation}
where $L(a_1,a_0|\mbox{data})$ is the likelihood function, $f(x)=[1-\exp(-x)]^{-1}$ is the Fermi function, and $k$ is the number of samples. The surface area of the subsystems in our quasi-one-dimensional model is constant, so we expect the small system values for $a_0$ and $a_1$ to vary linearly with $1/v$. Figure~\ref{fig:like} shows the linear fits to the coefficients obtained from the maximum likelihood analysis. The values of both $a_0$ and $a_1$ vary about 10\% of the exact result as $v$ goes from its smallest to largest value.
Extrapolating the fits to $1/v=0$ yields estimates of $a_0$ and $a_1$ within 0.5\% of the exact results.

\section{Discussion}
\label{sec:dis}
 
Our model provides an idealized representation of a granular material that samples the Edwards volume ensemble, which lets us examine the properties of a granular potential in a clear and rigorous way. In particular, we find that at constant volume, mass moves from high to low $\gamma/X$, until equilibrium is established when the total entropy of the system reaches its maximum, in analogy to the chemical potential in molecular systems.  Our work also shows that the entropic effect of adding a particle to the system changes sign as the compactivity because of the competing contributions from the partial molar volumes and entropy. This is likely to be quite general given the genetic evolution of thermodynamic parameters, $V/NX$ and $S/N$, observed in Fig.~\ref{fig:muox}. 

Bertin {\it et al.}~\cite{Bertain_Definition_2006} showed that intensive thermodynamic parameters can be defined in non-equilibrium systems, even in the absence of detailed balance, when the conditional steady state probability of finding a subsystem in a given state satisfies an asymptotic factorization. The factorization assumption becomes exact for one-dimensional systems that can be described by a transfer matrix, which is the case for our model. Thus, we provide a specific demonstration of a more general principle describing intensive thermodynamic potentials in this simple type of non-equilibrium system. However, it is important to note that the features of the granular potential observed here should be applicable to experimental systems in two- and three dimensions that have been shown to follow the Edwards volume ensemble~\cite{Yuan_Experimental_2021,Zeng_Equivalence_2022}, despite the simplicity of our model.

Equation~\ref{eq:gox} identifies the need for partial molar volumes and entropies that again have analogous functions to those observed in molecular systems, and play important roles in our understanding of mixtures. Chang {\it et al.}~\cite{Chang_Compaction_2022} recently took a step in this direction, using partial molar volumes and activity coefficients to describe jamming densities in bi-dispersed granular packings, and Liu {\it et al.}~\cite{Liu_Jammed_2022} suggest that a granular potential needs to be developed to understand binary mixtures of confined discs.  The granular potential could also play an important role in understanding phase equilibria in granular materials, where we expect both the compactivity and granular potentials to be equal in both phases at coexistence. For example, hard sphere packings appear to exhibit a first order freezing transition near random close packing from the amorphous packing to the face centred crystal~\cite{Jin_First_2010,Hanifpour_Mechanical_2014,Hanifour_Structural_2015}, as do sheared granular systems~\cite{Rietz_Nucleation_2018,Swinney_Homogeneous_2020}. Orientational ordering has been observed in the packings of cylindrical granular particles~\cite{Ding_Cubatic_2022}. 

Our example of equilibrium involving discs confined to two channels with different heights provides a limiting case for such a phase transition. The particles in the narrower channel occupy more space than those in the wider channel, even though they have the same two types of local packing environments, which gives rise to different volume distributions for the jammed packings in each channel. This mimics the distinct distributions we would expect to see in two phases of a granular system in higher dimensions. The coexistence in our model system occurs at $X=\infty$ because both distributions have the same entropy maximum, just located at different volumes, which gives $\Delta S=S_1-S_2=0$ and $\Delta V=V_1-V_2 >0$. However, in general, we would expect the high density phase of any two phase granular system to have a smaller number of jammed states and a lower entropy, which would move the coexistence point to lower $X$ and the transition would become first order with $\Delta S>0$ and $\Delta V>0$.

Finally, we demonstrated the applicability of a small system method for the calculation of the difference in $\gamma/X$ between two states when the bulk system is sampling the Edwards canonical ensemble by comparing our simulation results to the exact model analysis. The application of the method to systems in higher dimensions is more challenging because the surface area effects become more important and particle volumes become correlated over mesoscopic length scales, leading to system size effects. The shape of the subsystem may also play a role in higher dimensions as this may introduce edge and corner effects not present in our model. Nevertheless, our work highlights the utility of treating subsystems with boundary effects in granular materials and the method can be used in simulation and experiment where the particle positions can be visualized.

However, work is still required to put the granular potential on the same fundamental footing as the chemical potential. Our current analysis relies on the validity of the Edwards postulate and identification of the compactivity as the appropriate temperature-like variable.  This is true in our model, recent simulations suggest it is also true for jammed hard spheres near the jamming point~\cite{Martiniani_Numerical_2017} and recent experimental studies find the compactivity provides a meaningful measure of the granular temperature, but it may not be generally valid~\cite{Puckett_Equilibrating_2013}. We will need to examine the role of the granular potential in other ensembles, such as the stress ensemble~\cite{Blumenfeld_Inter_2012,Bililign_Protocol_2019}, where additional thermodynamic parameters appear in the fundamental equation.

\section{Conclusions}
\label{sec:con}
To conclude, we have shown that mass transport towards equilibrium in a simple model of a granular system is driven by differences in the ratio of the granular potential to the compactivity, and that the system comes to equilibrium when both the compactivity and the granular potential are uniform. The model also shows that the granular potential plays a role in determining the equilibrium between systems with different packing distributions, as we would expect in a phase transition, suggesting that the granular potential may be a useful concept in understanding a broad range of granular phenomena.

\section*{Author Contributions}
All authors contributed equally to the conceptualization, writing, reviewing and editing of this manuscript.

\section*{Conflicts of interest}
There are no conflicts to declare.

\section*{Acknowledgements}
We would like to acknowledge NSERC grant RGPIN--2019--03970 (J. M. G and R. K. B) for financial support. Computational resources and support were provided by the Prairie DRI Group and the Digital Research Alliance of Canada.

\appendix
\section{Kinetic Monte Carlo Method}
\label{sec:kmc}
This appendix describes the kinetic Monte Carlo Scheme used in our simulations. The state of the system can be characterized as a spin model, assigning 
particles in the most dense and defect states a 0 or 1 respectively. A single KMC move involves randomly selecting a particle and randomly swapping its local environment with the particle to the left or right. For example, $\{10\}\rightarrow\{01\}$ effectively moves a defect to the right, while a swap $\{00\}\rightarrow\{00\}$ has no net effect on the state. If the move results in a stable configuration, with no neighbouring defects, the move is accepted. 

If the move generates an unstable environment, the defect pair is eliminated ($\{11\}\rightarrow\{00\}$). However, the total number of defects must remain fixed to ensure the system stays jammed at constant $V$. A defect pair is then inserted by randomly selecting a particle. If it is a zero, then an attempt is made to insert a second defect two sites to the left or right with equal probability,  ($\{00000\}\rightarrow\{01010\}$). If a stable configuration is created, then the move is accepted, otherwise it is rejected. This is repeated until two defects are successfully inserted, returning the number of defects to its original value. One unit of time in the simulation corresponds to $N$ KMC moves.

\section{Grand Micro Canonical Ensemble}
\label{sec:gmc}
Here, we take a standard approach in statistical mechanics~\cite{} to develop the partition function for a system with fixed $V$, and $\gamma/X$. Consider a system of fixed volume that can exchange particles with a particle reservoir. The composite system, consisting of the system of interest and the reservoir, is isolated with fixed $N$ and $V$, so can it be described by the micro-canonical Edwards ensemble, where all the jammed microstates have equal probability. With the system in a fixed microstate with volume $V_s$, containing $N_s$ particles, the reservoir has $N_r=N-N_s$ particles, where $N_r>>N_s$, and volume $V_r=V-V_s$ where $V_r>>V_s$. However, $V_s$ is still sufficiently macroscopic that effects of the boundary between the reservoir and the system can be ignored.

The probability of finding the system in a microstate $s$ is given by,
\begin{equation}
P_s=\frac{1\times \Omega_r(N-N_s)}{\sum_s \Omega_r(N-N_s)}\mbox{,}\\
\label{eq:ps}
\end{equation}
where $\Omega_r(N-N_s)$ is the number of microstates in the reservoir. Taking the logarithm of Eq.~\ref{eq:ps} and expanding in terms of $N_s$ yields,
\begin{equation}
\begin{split}
\ln P_s&\approx C+\ln \Omega_r(N)-N_s\left(\frac{\partial \ln \Omega_r(N_r)}{\partial N_r}\right)_{N_r=N}\\
&=C+\ln \Omega_r(N)+\frac{N_s\gamma}{\lambda X}\mbox{,}\\
\end{split}
\label{eq:lnps}
\end{equation}
where $C$ is related to the denominator in Eq.~\ref{eq:ps} and we have used Eqs.~\ref{eq:entropy} and \ref{eq:dsdn} to connect $P_s$ to the granular potential imposed by the reservoir. Normalizing the probability then gives,
\begin{equation}
P_s=\frac{\exp[N_s \gamma/\lambda X]}{Z_{\gamma/X,V}}\mbox{,}\\
\label{eq:psZ}
\end{equation}
where, 
\begin{equation}
Z_{\gamma/X,V_s}=\sum_s \exp[N_s \gamma/\lambda X]=\sum_{N_s}\Omega(N_s,V_s)\exp[N_s \gamma/\lambda X]\mbox{,}\\
\label{eq:z}
\end{equation}
is the constant $\gamma/X,V_s$ partition function and the sum on the right is formed by collecting together all jammed microstates with the same number of particles contained in $V_s$. 

Finally, the partition function is connected directly to a thermodynamic potential through the Gibbs expression for the entropy, $S=-\lambda\sum_s P_s \ln P_s$, which gives,
\begin{equation}
\lambda X \ln Z_{\gamma/X,V_s}= XS+\gamma N_s=V_s\mbox{.}\\
\label{eq:zv}
\end{equation}

\section{Canonical Ensemble Simulation Method}
\label{sec:ces}
This appendix describes the canonical Monte Carlo simulation used to study the granular potential in our quasi-one-dimensional model. The state of the system of $N$ particles can be characterized using a spin model, assigning particles at the top and bottom of the channel as $1$ and $-1$ respectively. Note this differs from the model characterization used for the KMC. Two neighbouring particles with the same sign represent a defect state ($\{1,1\}$ and $\{-1,-1\}$), two successive particles with opposite signs represent a dense state ($\{1,-1\}$ and $\{-1,1\}$). An MC move consists of randomly selecting particle $i$ and attempting to swap signs with the $i+1$ neighbour. If the move generates an unstable configuration (pair of defects) the move is rejected. If the move results in a decrease in volume, it is immediately accepted. Moves that increase the volume $\Delta V$ are accepted with the probability $\exp[-\Delta V/\lambda X]$. An MC cycle consists of $N$ MC moves. The system is equilibrated for $500N$ MC cycles at each $X$.


\balance




\providecommand{\latin}[1]{#1}
\makeatletter
\providecommand{\doi}
  {\begingroup\let\do\@makeother\dospecials
  \catcode`\{=1 \catcode`\}=2 \doi@aux}
\providecommand{\doi@aux}[1]{\endgroup\texttt{#1}}
\makeatother
\providecommand*\mcitethebibliography{\thebibliography}
\csname @ifundefined\endcsname{endmcitethebibliography}
  {\let\endmcitethebibliography\endthebibliography}{}

\end{document}